\documentclass[iop]{emulateapj}

\shorttitle{ASTEROSEISMOLOGY OF 16 CYG A \& B}
\shortauthors{METCALFE ET AL.}

\begin{document}

\title{Asteroseismology of the solar analogs 16 Cyg A \& B from 
\textit{Kepler} observations}

\author{T.~S.~Metcalfe\altaffilmark{1,2,3},
W.~J.~Chaplin\altaffilmark{4,3},
T.~Appourchaux\altaffilmark{5},
R.~A.~Garc{\'i}a\altaffilmark{6,3},
S.~Basu\altaffilmark{7,3},
I.~Brand{\~a}o\altaffilmark{8},
O.~L.~Creevey\altaffilmark{9},
S.~Deheuvels\altaffilmark{7,3},
G.~Do\u{g}an\altaffilmark{1,3},
P.~Eggenberger\altaffilmark{10},
C.~Karoff\altaffilmark{11},
A.~Miglio\altaffilmark{4,3},
D.~Stello\altaffilmark{12},
M.~Y{\i}ld{\i}z\altaffilmark{13},
Z.~\c{C}elik\altaffilmark{13},
H.~M.~Antia\altaffilmark{14},
O.~Benomar\altaffilmark{12},
R.~Howe\altaffilmark{4},
C.~R{\'e}gulo\altaffilmark{15,16},
D.~Salabert\altaffilmark{9},
T.~Stahn\altaffilmark{17,18},
T.~R.~Bedding\altaffilmark{12,3},
G.~R.~Davies\altaffilmark{6},
Y.~Elsworth\altaffilmark{4},
L.~Gizon\altaffilmark{17,18},
S.~Hekker\altaffilmark{19,4},
S.~Mathur\altaffilmark{1,3},
B.~Mosser\altaffilmark{20},
S.~T.~Bryson\altaffilmark{21},
M.~D.~Still\altaffilmark{21},
J.~Christensen-Dalsgaard\altaffilmark{11,1,3},
R.~L.~Gilliland\altaffilmark{22},
S.~D.~Kawaler\altaffilmark{23,3},
H.~Kjeldsen\altaffilmark{11,3},
K.~A.~Ibrahim\altaffilmark{24},
T.~C.~Klaus\altaffilmark{24},
J.~Li\altaffilmark{25}}

\altaffiltext{1}{High Altitude Observatory, NCAR, P.O. Box 3000, Boulder CO 80307, USA}
\altaffiltext{2}{Computational \& Information Systems Laboratory, NCAR, P.O. Box 3000, Boulder CO 80307, USA}
\altaffiltext{3}{Kavli Institute for Theoretical Physics, Kohn Hall, University of California, Santa Barbara, CA 93106, USA}
\altaffiltext{4}{School of Physics and Astronomy, University of Birmingham, Edgbaston, Birmingham B15 2TT, United Kingdom}
\altaffiltext{5}{Univ Paris-Sud, Institut d'Astrophysique Spatiale, UMR8617, CNRS, B\^atiment 121, 91405 Orsay Cedex, France}
\altaffiltext{6}{Laboratoire AIM, CEA/DSM-CNRS-Universit\'e Paris Diderot; IRFU/SAp, Centre de Saclay, 91191 Gif-sur-Yvette Cedex, France}
\altaffiltext{7}{Department of Astronomy, Yale University, PO Box 208101, New Have, CT 06520-8101}
\altaffiltext{8}{Centro de Astrof\'{\i}sica and Faculdade de Ci\^encias, Universidade do Porto, Rua das Estrelas, 4150-762 Porto, Portugal}
\altaffiltext{9}{Laboratoire Lagrange, UMR7293, Universit\'e de Nice Sophia-Antipolis, CNRS, Observatoire de la C\^ote d'Azur, BP 4229, 06304 Nice Cedex 4, France}
\altaffiltext{10}{Observatoire de Gen\`eve, Universit\'e de Gen\`eve, 51 Ch. des Maillettes, CH-1290 Sauverny, Switzerland}
\altaffiltext{11}{Department of Physics and Astronomy, Aarhus University, DK-8000 Aarhus C, Denmark}
\altaffiltext{12}{Sydney Institute for Astronomy (SIfA), School of Physics, University of Sydney, NSW 2006, Australia}
\altaffiltext{13}{Ege University, Department of Astronomy and Space Sciences, Bornova, 35100, Izmir, Turkey}
\altaffiltext{14}{Remaining affiliations removed due to arXiv error}

\begin{abstract}

The evolved solar-type stars 16~Cyg~A \& B have long been studied as solar 
analogs, yielding a glimpse into the future of our own Sun. The orbital 
period of the binary system is too long to provide meaningful dynamical 
constraints on the stellar properties, but asteroseismology can help 
because the stars are among the brightest in the {\it Kepler} field. We 
present an analysis of three months of nearly uninterrupted photometry of 
16~Cyg~A \& B from the {\it Kepler} space telescope. We extract a total of 
46 and 41 oscillation frequencies for the two components respectively, 
including a clear detection of octupole ($l$=3) modes in both stars. We 
derive the properties of each star independently using the Asteroseismic 
Modeling Portal, fitting the individual oscillation frequencies and other 
observational constraints simultaneously. We evaluate the systematic 
uncertainties from an ensemble of results generated by a variety of 
stellar evolution codes and fitting methods. The optimal models derived by 
fitting each component individually yield a common age ($t=6.8\pm0.4$~Gyr) 
and initial composition ($Z_{\rm i}=0.024\pm0.002, Y_{\rm i}=0.25\pm0.01$) 
within the uncertainties, as expected for the components of a binary 
system, bolstering our confidence in the reliability of asteroseismic 
techniques. The longer data sets that will ultimately become available 
will allow future studies of differential rotation, convection zone 
depths, and long-term changes due to stellar activity cycles.

\end{abstract}


\slugcomment{The Astrophysical Journal (submitted)}

\section{INTRODUCTION}\label{sec1}

  \begin{figure*} 
  \epsscale{1.1}
  \plottwo{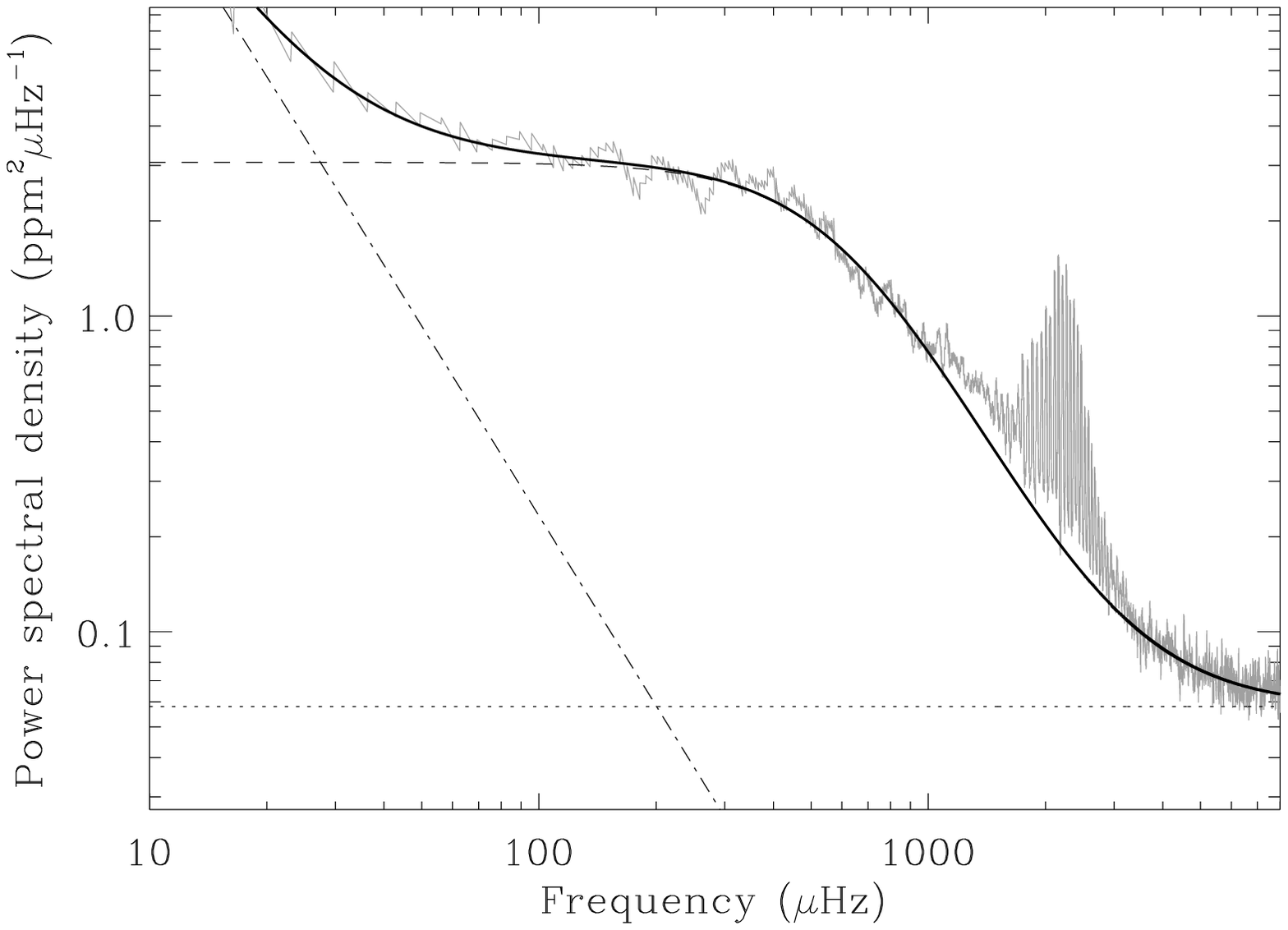}{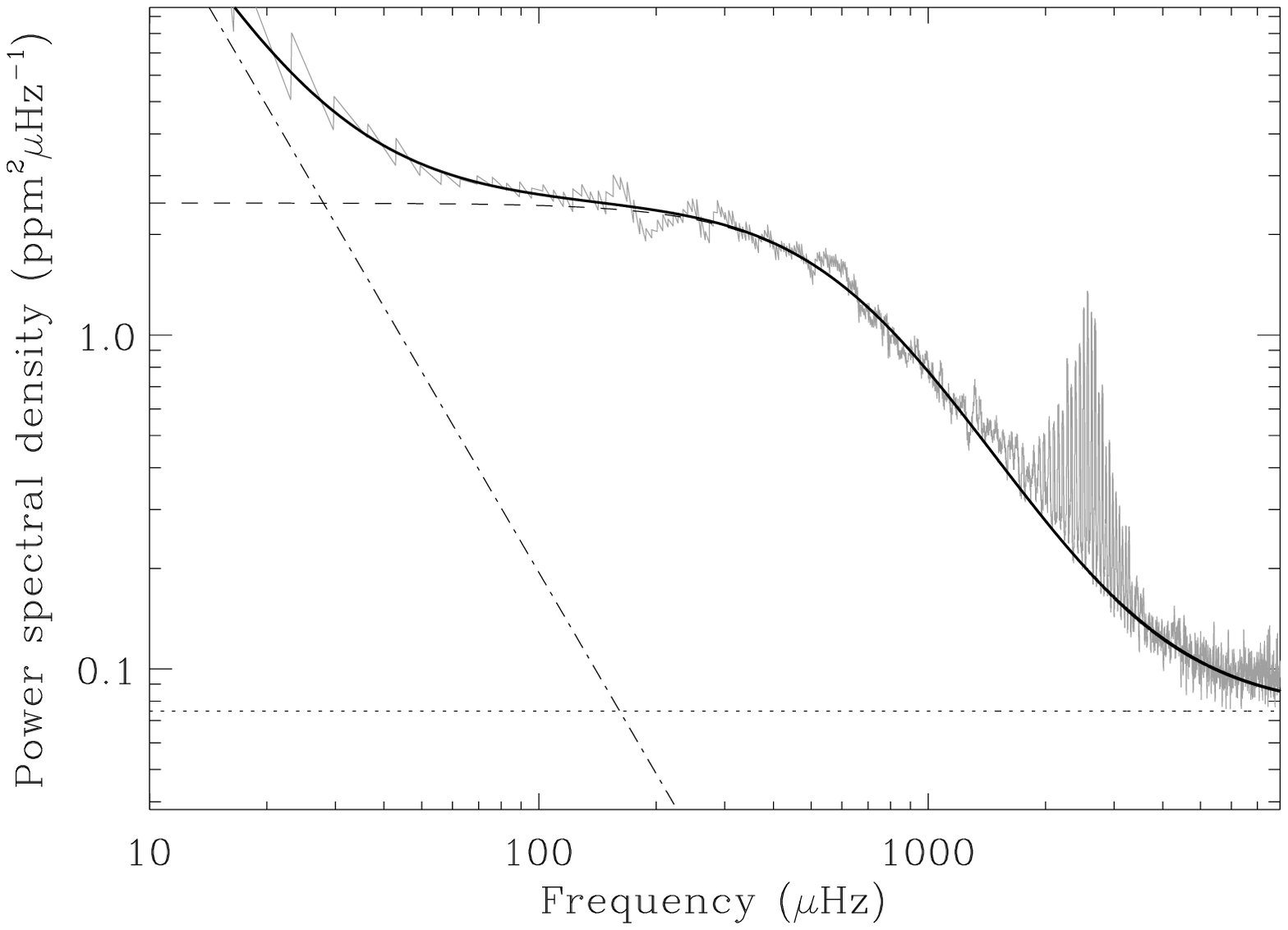} 
  \plottwo{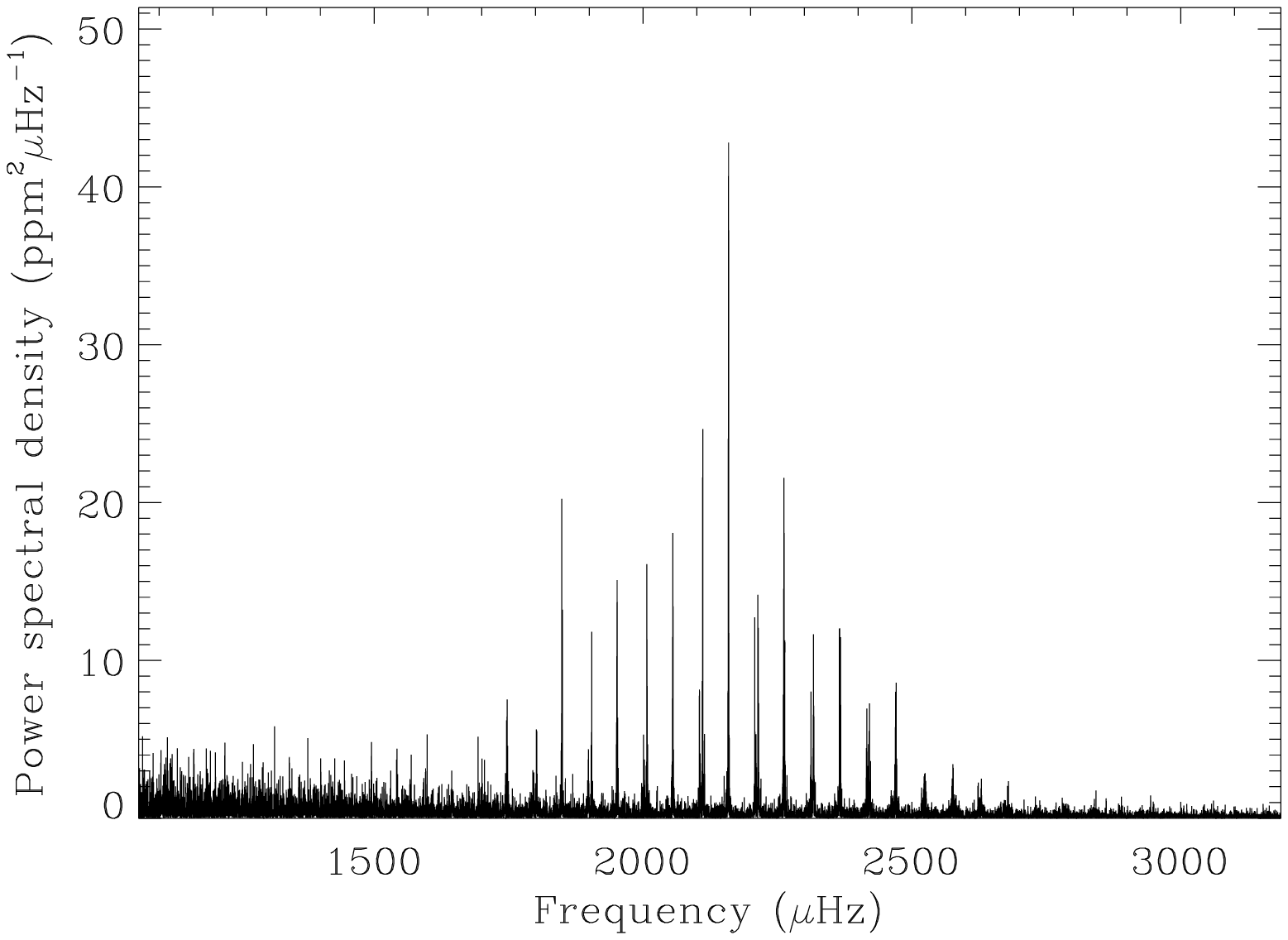}{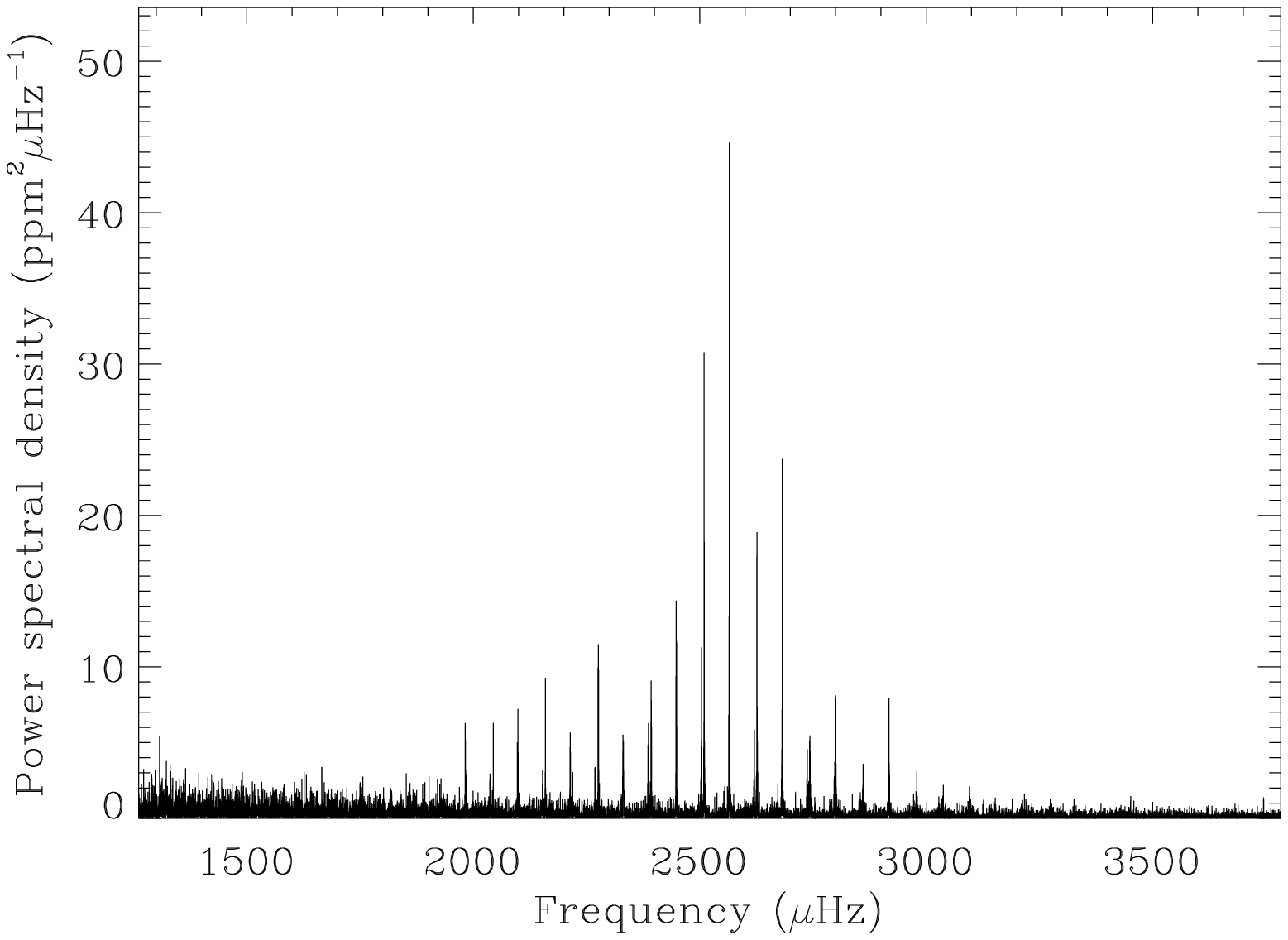}
  \caption{Power spectra of 16~Cyg~A (left panels) and 16~Cyg~B 
  (right panels). Top panels: 20~$\mu$Hz boxcar smoothed spectra (grey), 
  with best-fitting background components attributed to granulation (dashed 
  lines), stellar activity and/or larger scales of granulation (dot-dashed 
  lines) and shot noise (dotted lines), with the sum of the background 
  components plotted as solid black lines. Bottom panels: 
  Background-subtracted power spectra over the ranges in frequency where 
  high-order p modes are observed.\label{fig1}}
  \end{figure*}

Observations of the Sun provide an extraordinarily detailed snapshot of 
stellar structure and dynamics for a single set of physical properties and 
at a particular evolutionary state in the life of a star. To generalize 
our understanding of stellar evolution, and to evaluate the degree to 
which the Sun is typical or peculiar, it is useful to examine other 
classes of stars. \cite{cds96} defined several such classes, including: 
{\it solar twins} with fundamental physical properties very similar or 
identical to the Sun; {\it solar analogs} which are broadly comparable to 
the recent past and near future of the Sun; and {\it solar-like stars} 
including a wider range of F and G dwarfs and subgiants. Well-known solar 
twins such as 18~Sco \citep{baz11} provide some context for the Sun 
observed as a star, while solar analogs like $\kappa^1$~Cet, $\beta$~Hyi, 
and $\alpha$~Cen A \& B \citep{wal07, bra11, bed04, kje05} help calibrate 
stellar evolution for stars that are younger, older, and more or less 
massive than the Sun. Broader studies of solar-like stars probe the full 
range of relevant stellar properties and evolutionary states 
\citep[e.g.][]{cha11a,sil11}.

The bright stars 16~Cyg~A \& B ($\equiv$ HD~186408 \& 186427 $\equiv$ 
KIC~12069424 \& 12069449; V$\sim$6) have long been studied as solar 
analogs, with estimated ages near 6--8~Gyr \citep{wri04,vf05}. Although 
they are members of a hierarchical triple system with a red dwarf 
companion that is 10 magnitudes fainter \citep{tur01,pat02}, there are no 
dynamical constraints on the masses because the available data suggest an 
orbital period longer than 18,000 years \citep{hb99}. After the discovery 
of a 1.5 Jupiter-mass exoplanet in an eccentric 800-day orbit around 
16~Cyg~B \citep{coc97}, the system generated even more interest. Since 
then, both components have been monitored for magnetic activity, showing 
long-term variations around a mean chromospheric activity level well below 
that of the Sun at solar minimum (J.~Hall, private communication). So far 
there have been no direct measurements of rotation, but gyrochronology 
suggests that the rotation periods should be near 30 days \citep{sku72}. 
As two of the brightest stars in the {\it Kepler} field of view, 16~Cyg~A 
\& B can now be subjected to a long-term study that promises to yield more 
detailed information than is currently available for any star but the Sun.

  \begin{table*}
  \begin{center}
  \caption{Observed oscillation frequencies for 16 Cyg A \& B.\label{tab1}}
  \scriptsize
  \begin{tabular}{lcccclcccc}	
  \tableline\tableline
   & \multicolumn{4}{c}{16 Cyg A} & & \multicolumn{4}{c}{16 Cyg B} \\	
  \cline{2-5} \cline{7-10} \\	
  $n\tablenotemark{a}$ & $\ell=0\ (\mu$Hz) & $\ell=1\ (\mu$Hz) & $\ell=2\ (\mu$Hz) & $\ell=3\ (\mu$Hz) & &	
  $\ell=0\ (\mu$Hz) & $\ell=1\ (\mu$Hz) & $\ell=2\ (\mu$Hz) & $\ell=3\ (\mu$Hz) \\	
  \tableline
  13 &        \nodata     &        \nodata     & $1591.21 \pm 0.86$ &        \nodata     &&        \nodata     &        \nodata     &        \nodata     &        \nodata     \\
  14 & $1598.51 \pm 0.27$ & $1644.24 \pm 0.33$ & $1693.73 \pm 0.46$ & $1736.03 \pm 1.84$ &&        \nodata     &        \nodata     & $1920.99 \pm 0.24$ &        \nodata     \\
  15 & $1700.43 \pm 0.34$ & $1746.93 \pm 0.24$ & $1795.87 \pm 0.40$ & $1839.07 \pm 1.64$ && $1928.81 \pm 0.28$ & $1982.66 \pm 0.16$ & $2036.59 \pm 0.20$ &        \nodata     \\
  16 & $1802.15 \pm 0.17$ & $1849.11 \pm 0.13$ & $1898.08 \pm 0.27$ & $1944.07 \pm 1.57$ && $2044.21 \pm 0.15$ & $2098.20 \pm 0.17$ & $2152.91 \pm 0.19$ & $2202.75 \pm 0.65$ \\
  17 & $1904.62 \pm 0.15$ & $1951.98 \pm 0.16$ & $2001.82 \pm 0.17$ & $2045.09 \pm 0.80$ && $2159.36 \pm 0.16$ & $2214.00 \pm 0.18$ & $2269.07 \pm 0.21$ & $2317.08 \pm 0.44$ \\
  18 & $2007.45 \pm 0.13$ & $2055.41 \pm 0.16$ & $2105.60 \pm 0.15$ & $2150.15 \pm 0.19$ && $2276.03 \pm 0.12$ & $2330.88 \pm 0.16$ & $2386.30 \pm 0.17$ & $2436.78 \pm 0.33$ \\
  19 & $2110.94 \pm 0.11$ & $2158.89 \pm 0.12$ & $2208.90 \pm 0.19$ & $2253.41 \pm 0.35$ && $2392.87 \pm 0.14$ & $2448.17 \pm 0.11$ & $2503.56 \pm 0.13$ & $2553.00 \pm 0.23$ \\
  20 & $2214.33 \pm 0.17$ & $2262.32 \pm 0.16$ & $2312.49 \pm 0.29$ & $2356.92 \pm 0.46$ && $2509.75 \pm 0.13$ & $2565.35 \pm 0.10$ & $2619.99 \pm 0.23$ & $2672.34 \pm 0.28$ \\
  21 & $2317.18 \pm 0.17$ & $2366.15 \pm 0.16$ & $2416.24 \pm 0.33$ & $2461.26 \pm 1.04$ && $2626.43 \pm 0.11$ & $2682.38 \pm 0.14$ & $2737.44 \pm 0.31$ & $2788.74 \pm 1.40$ \\
  22 & $2420.75 \pm 0.30$ & $2470.23 \pm 0.25$ & $2520.91 \pm 0.81$ &        \nodata     && $2743.15 \pm 0.25$ & $2799.67 \pm 0.22$ & $2854.52 \pm 0.39$ & $2906.96 \pm 0.93$ \\
  23 & $2524.94 \pm 0.39$ & $2575.97 \pm 0.31$ & $2624.05 \pm 0.51$ &        \nodata     && $2860.63 \pm 0.26$ & $2917.75 \pm 0.22$ & $2972.73 \pm 0.70$ &        \nodata     \\
  24 & $2629.36 \pm 0.36$ & $2678.47 \pm 0.47$ & $2730.06 \pm 1.03$ &        \nodata     && $2978.95 \pm 0.40$ &        \nodata     & $3089.46 \pm 0.87$ &        \nodata     \\
  25 & $2736.22 \pm 1.45$ & $2783.71 \pm 1.22$ &        \nodata     &        \nodata     && $3096.00 \pm 0.54$ & $3152.45 \pm 0.61$ &        \nodata     &        \nodata     \\
  26 & $2838.68 \pm 0.38$ & $2889.61 \pm 0.38$ &        \nodata     &        \nodata     && $3215.94 \pm 0.91$ & $3274.63 \pm 0.55$ &        \nodata     &        \nodata     \\
  \tableline
  \end{tabular}
  \begin{minipage}{\textwidth}
  \tablenotetext{1}{Radial order $n$ from the optimal AMP models.}
  \end{minipage}
  \end{center}
  \end{table*}

In this Letter, we perform an asteroseismic analysis of the first three 
months of data on 16~Cyg~A \& B from the {\it Kepler} mission 
\citep{koc10}. Using the unprecedented observations, we model the two 
components independently and determine an identical age and initial 
composition, as expected for the members of a binary system. In 
\S\ref{sec2} we describe the data analysis methods, and in \S\ref{sec3} we 
present the asteroseismic modeling including an evaluation of both the 
statistical and systematic uncertainties. We conclude in \S\ref{sec4} with 
a discussion of the results and the potential for future studies utilizing 
the longer data sets that will soon become available.

\section{DATA ANALYSIS}\label{sec2}

The first full quarter of short-cadence observations \citep[58.85\,s 
sampling;][]{gil10} of 16~Cyg A \& B were obtained by {\it Kepler} between 
September and December 2010 (Q7). Both stars are significantly brighter 
than the photometric saturation limit. Saturated flux is conserved on {\it 
Kepler}, so no photometric precision is lost for saturated targets as long 
as the saturated pixels are included in the pixel aperture. Standard {\it 
Kepler} pixel apertures were not designed for such bright, saturated 
targets and in the case of 16~Cyg contain a prohibitively large number of 
unneeded pixels. Custom masks were therefore defined from Q3 full frame 
images to capture all of the flux using fewer pixels. The raw photometric 
light-curves extracted from these masks \citep{jen10} were then prepared 
for asteroseismic analysis in the manner described by \cite{gar11}. 
Figure~\ref{fig1} shows the power spectra of both stars (16~Cyg~A in the 
left panels, and 16~Cyg~B in the right panels).

The top panels include boxcar smoothed power spectra (in grey) over an 
extended range in frequency, showing not only the Gaussian-like power 
excess due to solar-like oscillations, but also contributions to the 
background power-spectral density attributable to granulation (dashed 
lines), stellar activity and/or larger scales of granulation (dot-dashed 
lines) and shot noise (dotted lines). The backgrounds were fit with a 
three-component model, comprising two Harvey-like power laws \citep{har85} 
to represent granulation and activity, and a flat component to represent 
the contribution of shot noise.  The best-fitting sum of background 
components is shown as a solid black line in each panel. We note that the 
best-fitting timescales ($\tau_{\rm gran,A}=257 \pm 6$~sec, $\tau_{\rm 
gran,B}=241 \pm 8$~sec) and peak powers ($P_{\rm gran,A}=3.01 \pm 
0.08$~ppm$^2/\mu$Hz, $P_{\rm gran,B}=2.41 \pm 0.07$~ppm$^2/\mu$Hz) of the 
granulation components are both slightly greater than the solar values 
estimated from Sun-as-a-star observations, and follow the scaling 
relations derived by \cite{kb11}.

The bottom panels show very clear patterns of peaks due to solar-like 
oscillations of high radial order, $n$. The quality of the oscillation 
spectra are exquisite, with each star showing more than fifteen radial 
overtones including many octupole ($l$=3) modes. The maximum peak 
height-to-background ratios are comparable to those observed in 
photometric Sun-as-a-star data---the shot noise level is so low in these 
{\it Kepler} data that the intrinsic stellar granulation noise dominates 
the background across the frequency ranges where the modes are observed.

Ten teams provided estimates of the frequencies of the observed modes, 
applying {\it peak-bagging} techniques developed for application to CoRoT 
\citep{app08} and {\it Kepler} data \citep[e.g., see][]{cam11,mat11}. 
These techniques varied in the details of the optimizations 
performed---which included classical maximum-likelihood estimation 
\citep[e.g.][]{fle11} and Markov Chain Monte Carlo methods 
\citep[e.g.][]{hc11}---and in the number of free parameters and 
assumptions made for fits of Lorentzian-like models to mode peaks in the 
power spectra.

The results of the ten teams were analyzed to produce final frequency sets 
for each star. First, we sought to identify objectively those modes for 
which a robust, well-determined frequency could be estimated. This 
involved two types of checks. In one, we identified a list of modes with 
good agreement between the best-fitting frequencies from various teams. 
This was achieved using a modified version of the procedure outlined in 
\cite{cam11} and \cite{mat11}. A so-called {\it minimal frequency set} of 
modes was produced, for which a majority of the teams' estimates were 
retained after applying Peirce's criterion \citep{pei52,gou55} for outlier 
rejection. A second set of checks involved visual inspection of the 
frequency-power spectra (and \'echelle diagrams of those spectra), 
combined with objective false alarm probability \citep[e.g.][]{cha02} and 
likelihood ratio tests \citep[e.g.][]{app11}.

  \begin{figure*}
  \epsscale{1.05}
  \plottwo{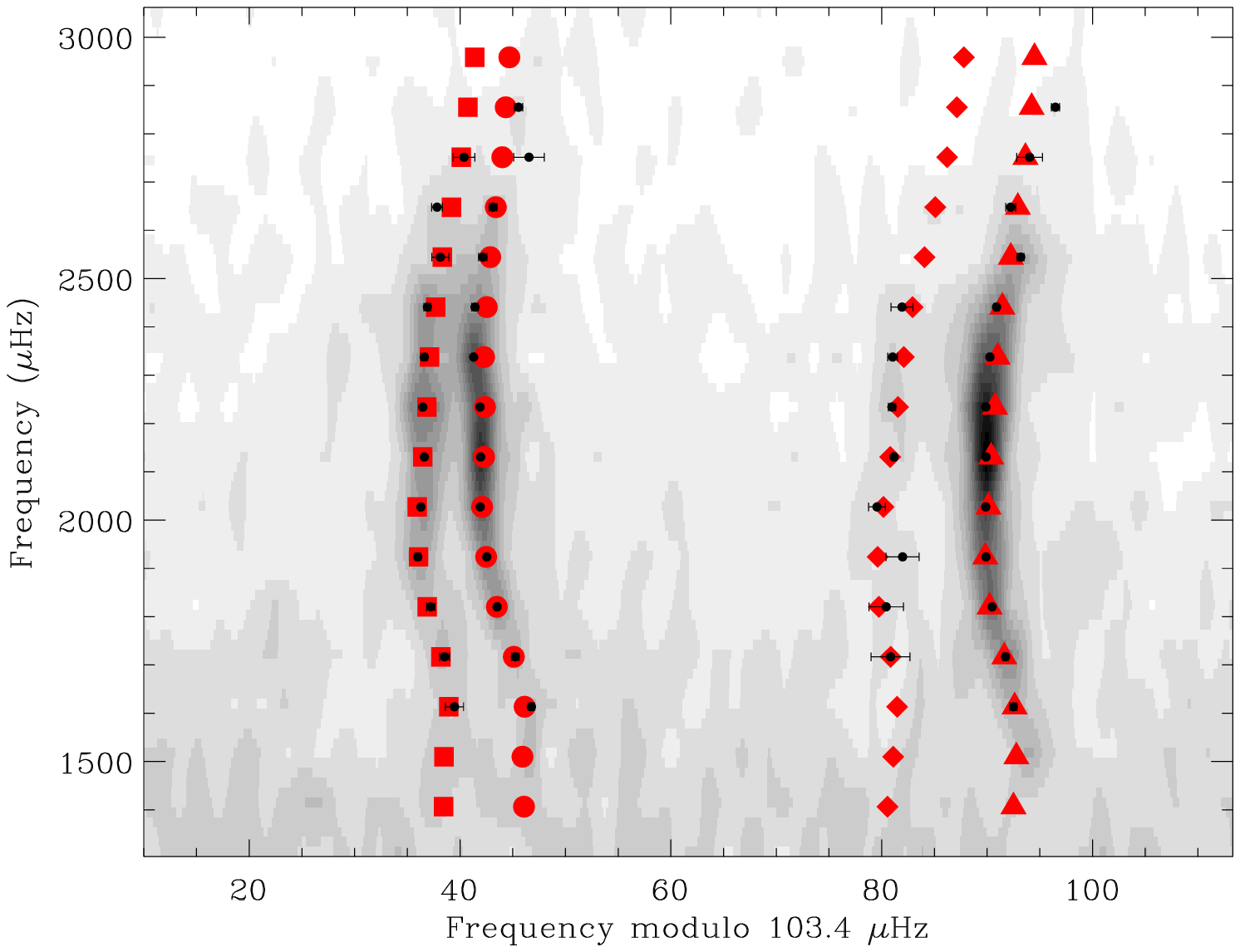}{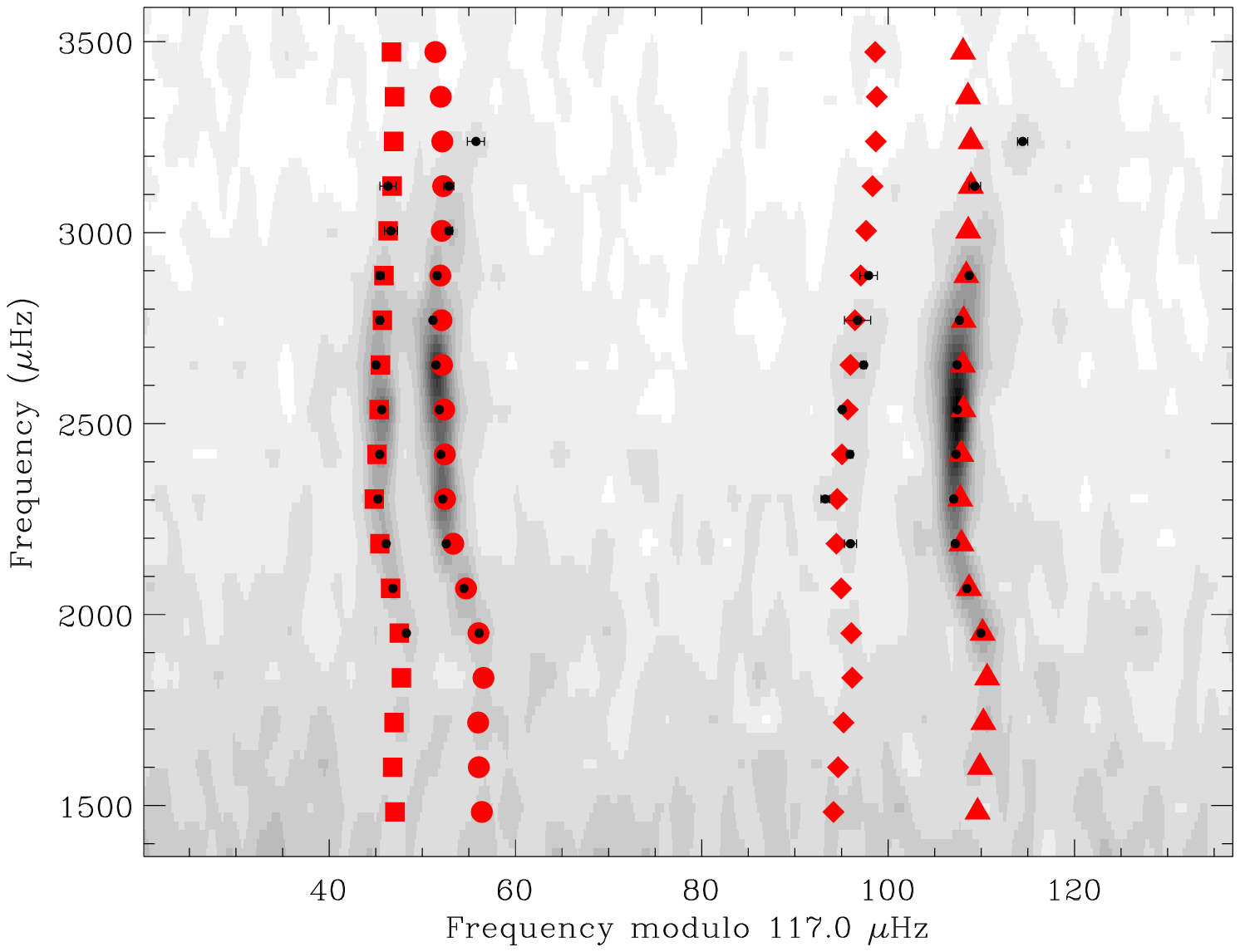}
  \caption{\'Echelle diagrams of 16\,Cyg~A (left) and 16\,Cyg~B (right), 
  showing the observed frequencies as black points with horizontal error 
  bars. The frequencies of the optimal models from AMP are shown using 
  different red symbols to indicate modes with radial ($l$=0, circles), 
  dipole ($l$=1, triangles), quadrupole ($l$=2, squares) and octupole 
  ($l$=3, diamonds) geometry. A greyscale map showing a Gaussian smoothed 
  power spectrum (FWHM $\sim2~\mu$Hz) is included in the background for 
  reference.\label{fig2}}
  \end{figure*}

With a list of robust modes in hand, one of the teams was then selected to 
re-fit these modes in both stars using a single Lorentzian profile per 
mode (i.e.\ no rotational splitting and the inclination angle fixed at 
0$^\circ$). This team was chosen as the one whose initial best-fitting 
frequencies showed the closest match to the frequencies of the 
minimal-set. Use of frequencies from one of the teams, as opposed to some 
average over all teams, meant that the modeling could rely on an easily 
reproducible set of input frequencies (see Table~\ref{tab1}).

\section{ASTEROSEISMIC MODELING}\label{sec3}

The set of oscillation modes from the peak-bagging analysis described in 
\S\ref{sec2} included a total of 46 and 41 individual frequencies for 
16~Cyg~A \& B, respectively. As inputs for the stellar modeling, we 
supplemented these asteroseismic constraints with the spectroscopic 
properties of each component derived by \cite{ram09} [$T_{\rm 
eff,A}=5825\pm50$~K, $\log g_{\rm A}=4.33\pm0.07$, [Fe/H]$_{\rm 
A}=0.096\pm0.026$; $T_{\rm eff,B}=5750\pm50$~K, $\log g_{\rm 
B}=4.34\pm0.07$, [Fe/H]$_{\rm B}=0.052\pm0.021$]. Using these $T_{\rm 
eff}$ values to obtain bolometric corrections from \cite{flo96} and 
adopting $M_{\rm bol}=4.73\pm0.03$ from \cite{tor10}, we combined the 
extinction estimates from \cite{amm06} with the updated Hipparcos 
parallaxes \citep{van07} to derive luminosity constraints: $L_A = 
1.56\pm0.05\ L_\odot$, $L_B = 1.27\pm0.04\ L_\odot$.

We calculated separate values of $\chi^2$ for the asteroseismic and 
spectroscopic constraints and attempted to minimize them simultaneously 
using the Asteroseismic Modeling Portal \citep[AMP;][]{met09}. This 
automated method uses a parallel genetic algorithm to search a broad range 
of stellar parameters and objectively determines the globally optimal 
model for a given set of observations. Although 16~Cyg~A \& B are members 
of a binary system and presumably formed simultaneously from the same 
material, we fit each set of constraints independently and did not force 
the models to have a common age or initial composition. The oscillation 
frequencies of the optimal models for each star are plotted as red symbols 
in Figure~\ref{fig2}, where the observed modes are shown as black points 
with horizontal error bars. In both cases, the asteroseismic $\chi^2$ is 
less than 10 and the spectroscopic $\chi^2$ is less than 1, so the models 
represent a reasonably good match to both sets of observational 
constraints. In Table~\ref{tab2} we list the optimal values for the radius 
($R$) and for the adjustable model parameters, including: the mass ($M$), 
age ($t$), initial metallicity ($Z_{\rm i}$) and helium mass fraction 
($Y_{\rm i}$), and the mixing-length parameter ($\alpha$), along with the 
asteroseismic $\chi^2$. The statistical uncertainties on each parameter 
($\sigma_{\rm stat}$) were determined using Singular Value Decomposition.

To evaluate the possible sources of systematic uncertainty from the 
ingredients and assumptions in our models, six teams were given the 
results from AMP and asked to reproduce the fit using the same set of 
observational constraints with their own stellar evolution codes and 
fitting methods. The physical ingredients adopted by each team differed 
slightly from those employed by AMP\footnote{AMP uses the OPAL 2005 
equation of state and the most recent OPAL opacities supplemented by 
\cite{af94} opacities at low temperature, nuclear reaction rates from 
\cite{bp92}, and includes the effects of helium diffusion and settling 
following \cite{mp93}. Convection is treated with standard mixing-length 
theory without overshooting \citep{bv58}. ANK\.I solves the Saha equation 
itself, uses low temperature opacities from \cite{fer05}, nuclear reaction 
rates primarily from \cite{cf88}, and includes a full treatment of 
diffusion following \cite{tbl94}. ASTEC1 and ASTEC2 use \cite{fer05} 
opacities, NACRE reaction rates \citep{ang99}, and neglect diffusion. 
CESAM uses NACRE reaction rates and treats convection following 
\cite{cgm96}. Geneva uses NACRE reaction rates, and treats diffusion 
following \cite{paq86}. YREC uses \cite{fer05} opacities, nuclear reaction 
rates primarily from \cite{ade98}, treats diffusion following \cite{tbl94} 
and includes convective overshooting. All models include the empirical 
correction for surface effects proposed by \cite{kbc08}.}, allowing us to 
explore the degree of model-dependence in our results. The optimal 
parameter values from each team are listed in Table~\ref{tab2}, where we 
combine all of them into an adopted value (bold row) representing the 
average of the individual estimates weighted by $1/\chi^2$. The systematic 
uncertainty ($\sigma_{\rm sys}$) on each parameter reflects the variance 
of the results, again weighted by $1/\chi^2$.

  \begin{table*}
  \begin{center}
  \caption{Stellar model-fitting results for 16 Cyg A \& B.\label{tab2}}
  \scriptsize
  \begin{tabular}{rccccccrcccccccr}
  \tableline\tableline
   & \multicolumn{7}{c}{16 Cyg A} & & \multicolumn{7}{c}{16 Cyg B} \\
  \cline{2-8} \cline{10-16} \\ 
   & $R/R_\odot$ & $M/M_\odot$ & $t$(Gyr) & $Z_{\rm i}$ & $Y_{\rm i}$ & $\alpha$ & $\chi^2$ & &
     $R/R_\odot$ & $M/M_\odot$ & $t$(Gyr) & $Z_{\rm i}$ & $Y_{\rm i}$ & $\alpha$ & $\chi^2$ \\
  \tableline
  AMP\dotfill & 1.236 & 1.10 & 6.5 & 0.022 & 0.25 & 2.06 &  5.47 && 1.123 & 1.06 & 5.8 & 0.020 & 0.25 & 2.05 & 9.80 \\
  $\sigma_{\rm stat}$ & 0.016 & 0.01 & 0.2 & 0.002 & 0.01 & 0.03 & \nodata && 0.020 & 0.01 & 0.1 & 0.001 & 0.01 & 0.03 & \nodata \\
   & & & & & & & & & & & & & & & \\
  ANK\.I\dotfill & 1.260 & 1.14 & 6.4 & 0.024 & 0.26 & 1.94 & 21.41 && 1.138 & 1.08 & 6.4 & 0.022 & 0.26 & 1.94 & 23.29 \\
  ASTEC1\dotfill & 1.237 & 1.10 & 7.5 & 0.023 & 0.25 & 2.00 &  5.70 && 1.121 & 1.05 & 7.3 & 0.021 & 0.25 & 2.00 &  7.97 \\
  ASTEC2\dotfill & 1.235 & 1.10 & 6.8 & 0.022 & 0.25 & 2.00 &  7.70 && 1.134 & 1.09 & 6.3 & 0.025 & 0.25 & 2.00 &  8.47 \\
  CESAM\dotfill& 1.253 & 1.14 & 7.0 & 0.027 & \phantom{\tablenotemark{a}}0.24\tablenotemark{a} & \phantom{\tablenotemark{b}}0.72\tablenotemark{b} &  3.53 && 1.136 & 1.09 & 6.9 & 0.025 & \phantom{\tablenotemark{a}}0.24\tablenotemark{a} & \phantom{\tablenotemark{b}}0.73\tablenotemark{b} &  4.78 \\
  Geneva\dotfill& 1.236 & 1.10 & \phantom{\tablenotemark{c}}6.7\tablenotemark{c} & \phantom{\tablenotemark{c}}0.024\tablenotemark{c} & \phantom{\tablenotemark{c}}0.26\tablenotemark{c} & \phantom{\tablenotemark{c}}1.80\tablenotemark{c} & 10.82 && 1.122 & 1.06 & \phantom{\tablenotemark{c}}6.7\tablenotemark{c} & \phantom{\tablenotemark{c}}0.024\tablenotemark{c} & \phantom{\tablenotemark{c}}0.26\tablenotemark{c} & \phantom{\tablenotemark{c}}1.80\tablenotemark{c} & 10.98 \\
  YREC\dotfill& 1.244 & 1.11 & 6.9 & 0.026 & 0.26 & 2.08 &  5.68 && 1.121 & 1.05 & \phantom{\tablenotemark{d}}6.9\tablenotemark{d} & 0.022 & 0.26 & 1.84 &  3.17 \\
   & & & & & & & & & & & & & & & \\
  ~~~{\bf adopted} & {\bf 1.243} & {\bf 1.11} & {\bf 6.9} & {\bf 0.024} & {\bf 0.25} & {\bf 2.00} & \nodata && {\bf 1.127} & {\bf 1.07} & {\bf 6.7} & {\bf 0.023} & {\bf 0.25} & {\bf 1.92} & \nodata \\
  $\sigma_{\rm sys}$ & 0.008 & 0.02 & 0.3 & 0.002 & 0.01 & 0.08 & \nodata && 0.007 & 0.02 & 0.4 & 0.002 & 0.01 & 0.09 & \nodata \\
  \tableline
  \end{tabular}
  \begin{minipage}{\textwidth}
  \tablenotetext{1}{Values of $Y_{\rm i}<0.24$ excluded from search.}
  \tablenotetext{2}{Value of $\alpha$ from the \cite{cgm96} treatment of convection, excluded from average.}
  \tablenotetext{3}{Age, composition, and mixing-length constrained to be identical in both components.}
  \tablenotetext{4}{Age of 16 Cyg B constrained to be identical to the value found for 16 Cyg A.}
  \end{minipage}
  \end{center}
  \end{table*}


As expected, there are slight differences between the optimal parameter 
values determined by each team. Since we effectively used AMP to solve the 
global optimization problem, these differences reflect subtle shifts in 
the locally optimal solution due to the physical ingredients included in 
each stellar evolution code. However, the results from different teams 
also include small offsets due to incomplete optimization---refined 
sampling of each adjustable parameter will always improve the fit, and 
there was no uniform criterion for when to stop fine tuning. To minimize 
the influence of this {\it technique error} on the final results, we 
weight the average parameter values and uncertainties using $1/\chi^2$ 
from each result as a proxy for the overall quality of the fit. This 
ensures that the variance reflects the actual systematic differences 
between model physics rather than the effort expended by each team in 
trying to match the observations. As with AMP, most of the teams did not 
force any of the model parameters of 16~Cyg~A \& B to share a common 
value. The exceptions were the Geneva code (which forced a common age, 
initial composition and mixing-length), and YREC (which forced the model 
for B to have the same age as the optimal model for A). Excluding these 
models from the average does not significantly change the values of the 
adopted parameters listed in Table~\ref{tab2}.

\section{RESULTS \& DISCUSSION}\label{sec4}

We have performed an analysis of the solar analogs 16~Cyg~A \& B using 
three months of observations from the {\it Kepler} space telescope, 
yielding the highest quality asteroseismic data sets for any star but the 
Sun (see Figure~\ref{fig1}). We identify a total of 46 and 41 oscillation 
frequencies in the two components respectively, including a clear 
detection of octupole ($l$=3) modes in both stars. These modes are 
difficult to detect in photometric data because the bright and dark 
patches associated with higher degree modes are normally expected to 
cancel in disk-integrated measurements. The unambiguous detection of such 
modes from the {\it Kepler} light-curves of 16~Cyg~A \& B is a testament 
to the exceptional quality of the data.

We derived the properties of each star independently by fitting stellar 
models to the oscillation frequencies (see Table~\ref{tab1}) and other 
observational constraints (see \S\ref{sec3}) simultaneously. The initial 
results from fitting each star individually using the Asteroseismic 
Modeling Portal (AMP; see Figure~\ref{fig2}) yield the same initial 
composition within the statistical uncertainties, and a similar age for 
the two components. Further analysis using several stellar evolution codes 
employing a variety of input physics allows us to quantify the 
model-dependence of our results, and to adopt reliable values and 
uncertainties from the ensemble. The adopted stellar properties of 
16~Cyg~A \& B (see Table~\ref{tab2}) reinforce the conclusion that the two 
stars share a common age ($t=6.8\pm0.4$~Gyr) and initial composition 
($Z_{\rm i}=0.024\pm0.002, Y_{\rm i}=0.25\pm0.01$), as expected for a 
binary system but without imposing this as a constraint for the modeling. 
This fundamental result bolsters our confidence in the reliability of 
asteroseismic inferences of stellar structure and evolution.

The relative size of the statistical and systematic uncertainties provides 
an important benchmark for what we can expect from asteroseismology with 
the {\it Kepler} mission. The statistical uncertainties on the stellar 
radii from AMP were derived from the distribution of radii in an ensemble 
of models that differ from the optimal model by $\pm1\sigma$ for each 
adjustable parameter. Such estimates implicitly include the influence of 
parameter correlations, and are consequently much larger than the 
systematic variation in optimal radii from different stellar evolution 
codes. This is not the case for adjustable model parameters like the mass 
and age, where $\sigma_{\rm sys}$ can be 2--4 times larger than 
$\sigma_{\rm stat}$, while the two are roughly comparable for the initial 
metallicity and helium mass fraction. The mixing-length parameter is a 
special case, because the range of estimates from different stellar 
evolution codes reflect variations in the solar-calibrated values that 
arise from differences in the input physics and in the specific 
formulation of mixing-length theory that is implemented in each code. 
Thus, the systematic uncertainties on $\alpha$ are likely to be 
overestimated, and small variations in the optimal value of this parameter 
from different codes should not be overinterpreted.

These extraordinary results were possible using just the first three 
months of short-cadence observations (Q7) from {\it Kepler}. Nine months 
of data will soon be available (Q7-8-9), and the stars continue to be on 
the short-cadence target list---at least through Q12, and hopefully for 
the remainder of the mission. These longer data sets will gradually yield 
higher frequency precision and improve the signal-to-noise ratio in the 
power spectra, enabling further characterization of the stars from the 
current frequency sets and facilitating detection of additional 
oscillation modes at higher and lower frequencies.

From 6--9 months of data we may begin to resolve rotational splitting of 
the non-radial oscillation modes into multiple components with different 
azimuthal order, $m$. The variation of this splitting as a function of the 
radial order $n$ can probe radial differential rotation, while the 
differences between non-radial modes with different spherical degree $l$ 
can reveal latitudinal variations. Such measurements of rotation may help 
to constrain possible scenarios to explain the different Li abundances of 
the two stars \citep{sch11,ram11}. With 12--18 months of data, the 
frequency precision may be sufficient to resolve oscillatory signals in 
the deviations from uniform frequency spacing (so-called second 
differences, $\delta_2\nu$) which reflect the acoustic depths of sharp 
transitions in the stellar structure, such as the helium ionization region 
and the base of the surface convection zone. Even longer data sets will 
allow us to probe the influence of stellar activity cycles, which lead to 
small anti-correlated changes in the frequencies and amplitudes of the 
oscillation modes. By the end of the baseline {\it Kepler} mission, and 
hopefully through an extended mission, these two bright solar analogs 
promise to yield the clearest picture yet of the future of our own Sun.


\acknowledgments Funding for this Discovery mission is provided by NASA's 
Science Mission Directorate. This work was supported in part by NASA grant 
NNX09AE59G. Computational time on Kraken at the National Institute of 
Computational Sciences was provided through NSF TeraGrid allocation 
TG-AST090107. We acknowledge the KITP staff at UCSB for their warm 
hospitality during the research program ``Asteroseismology in the Space 
Age''. This research was supported in part by the National Science 
Foundation under Grant No.\ NSF PHY05-51164. The authors would like to 
thank Jeff Hall, Todd Henry, Dave Soderblom, and Russel White for helpful 
discussions, as well as the entire Kepler team, without whom these results 
would not be possible. We also thank all funding councils and agencies 
that have supported the activities of the Kepler Asteroseismic Science 
Consortium Working Group 1, including the Pale Blue Dot Project hosted by 
White Dwarf Research Corporation ({\tt http://whitedwarf.org/palebluedot/}).



\begin{thebibliography}{}

\bibitem[Adelberger et al.(1998)]{ade98} Adelberger, E.~G., Austin, S.~M., 
Bahcall, J.~N., et al.\ 1998, Rev.~Mod.~Phys., 70, 1265

\bibitem[Alexander \& Ferguson(1994)]{af94} Alexander, D.~R., \& Ferguson, 
J.~W.\ 1994, \apj, 437, 879

\bibitem[Ammons et al.(2006)]{amm06} Ammons, S.~M., Robinson, S.~E., 
Strader, J., et al.\ 2006, \apj, 638, 1004

\bibitem[Angulo et al.(1999)]{ang99} Angulo, C., Arnould, M., Rayet, M., 
et al.\ 1999, Nuclear Physics A, 656, 3

\bibitem[Appourchaux(2011)]{app11} Appourchaux, T.\ 2011, Canary Islands 
Winter School of Astrophysics, Cambridge University Press, 
(arXiv:1103.5352).

\bibitem[Appourchaux et al.(2008)]{app08} Appourchaux, T., Michel, E., 
Auvergne, M., et al.\ 2008, \aap, 488, 705

\bibitem[Bahcall \& Pinsonneault(1992)]{bp92} Bahcall, J.~N., \& 
Pinsonneault, M.~H.\ 1992, Rev.~Mod.~Phys., 64, 885

\bibitem[Bazot et al.(2011)]{baz11} Bazot, M., Ireland, M.~J., Huber, D., 
et al.\ 2011, \aap, 526, L4

\bibitem[Bedding et al.(2004)]{bed04} Bedding, T.~R., Kjeldsen, H., 
Butler, R.~P., et al.\ 2004, \apj, 614, 380

\bibitem[B{\"o}hm-Vitense(1958)]{bv58} {B{\"o}hm-Vitense}, E. 1958, 
Zeitschrift fur Astrophysik, 46, 108

\bibitem[Brand{\~a}o et al.(2011)]{bra11} Brand{\~a}o, I.~M., Do{\u g}an, 
G., Christensen-Dalsgaard, J., et al.\ 2011, \aap, 527, A37

\bibitem[Campante et al.(2011)]{cam11} Campante, T.~L., Handberg, R., 
Mathur, S., et al.\ 2011, \aap, 534, A6

\bibitem[Canuto et al.(1996)]{cgm96} Canuto, V.~M., Goldman, I., \& 
Mazzitelli, I.\ 1996, \apj, 473, 550

\bibitem[Caughlan \& Fowler(1988)]{cf88} Caughlan, G.~R., \& Fowler, 
W.~A.\ 1988, Atomic Data and Nuclear Data Tables, 40, 283

\bibitem[Cayrel de Strobel(1996)]{cds96} Cayrel de Strobel, G.\ 1996, 
\aapr, 7, 243

\bibitem[Chaplin et al.(2002)]{cha02} Chaplin, W.~J., Elsworth, Y., Isaak, 
G.~R., et al.\ 2002, \mnras, 336, 979

\bibitem[Chaplin et al.(2011)]{cha11a} Chaplin, W.~J., Kjeldsen, H., 
Christensen-Dalsgaard, J., et al.\ 2011, Science, 332, 213

\bibitem[Cochran et al.(1997)]{coc97} Cochran, W.~D., Hatzes, A.~P., 
Butler, R.~P., \& Marcy, G.~W.\ 1997, \apj, 483, 457

\bibitem[Ferguson et al.(2005)]{fer05} Ferguson, J.~W., Alexander, D.~R., 
Allard, F., et al.\ 2005, \apj, 623, 585

\bibitem[Fletcher et al.(2011)]{fle11} Fletcher, S.~T., Broomhall, A.-M., 
Chaplin, W.~J., et al.\ 2011, \mnras, 413, 359

\bibitem[Flower(1996)]{flo96} Flower, P.~J.\ 1996, \apj, 469, 355

\bibitem[Garc{\'{\i}}a et al.(2011)]{gar11} Garc{\'{\i}}a, R.~A., Hekker, 
S., Stello, D., et al.\ 2011, \mnras, 414, L6

\bibitem[Gilliland et al.(2010)]{gil10} Gilliland, R.~L., Jenkins, J.~M., 
Borucki, W.~J., et al.\ 2010, \apjl, 713, L160

\bibitem[Gould(1855)]{gou55} Gould, B.~A.\ 1855, \aj, 4, 81

\bibitem[Handberg \& Campante(2011)]{hc11} Handberg, R., \& Campante, 
T.~L.\ 2011, \aap, 527, A56

\bibitem[Harvey(1985)]{har85} Harvey, J.\ 1985, Future Missions in Solar, 
Heliospheric \& Space Plasma Physics, 235, 199

\bibitem[Hauser \& Marcy(1999)]{hb99} Hauser, H.~M., \& Marcy, G.~W.\ 
1999, \pasp, 111, 321

\bibitem[Jenkins et al.(2010)]{jen10} Jenkins, J.~M., Caldwell, D.~A., 
Chandrasekaran, H., et al.\ 2010, \apjl, 713, L87

\bibitem[Kjeldsen et al.(2005)]{kje05} Kjeldsen, H., Bedding, T.~R., 
Butler, R.~P., et al.\ 2005, \apj, 635, 1281

\bibitem[Kjeldsen et al.(2008)]{kbc08} Kjeldsen, H., Bedding, T.~R., \& 
Christensen-Dalsgaard, J.\ 2008, \apjl, 683, L175

\bibitem[Kjeldsen \& Bedding(2011)]{kb11} Kjeldsen, H., \& Bedding, T.~R.\ 
2011, \aap, 529, L8

\bibitem[Koch et al.(2010)]{koc10} Koch, D.~G., Borucki, W.~J., Basri, G., 
et al.\ 2010, \apjl, 713, L79

\bibitem[Mathur et al.(2011)]{mat11} Mathur, S., Handberg, R., Campante, 
T.~L., et al.\ 2011, \apj, 733, 95

\bibitem[Metcalfe et al.(2009)]{met09} Metcalfe, T.~S., Creevey, O.~L., \& 
Christensen-Dalsgaard, J.\ 2009, \apj, 699, 373

\bibitem[Michaud \& Proffitt(1993)]{mp93} Michaud, G., \& Proffitt, C.~R.\ 
1993, in Proc.~IAU Colloq.~137: Inside the stars, eds A. Baglin, \& W.~W. 
Weiss, ASP Conf., 40, 246

\bibitem[Paquette et al.(1986)]{paq86} Paquette, C., Pelletier, C., 
Fontaine, G., \& Michaud, G.\ 1986, \apjs, 61, 177

\bibitem[Patience et al.(2002)]{pat02} Patience, J., White, R.~J., Ghez, 
A.~M., et al.\ 2002, \apj, 581, 654

\bibitem[Peirce(1852)]{pei52} Peirce, B.\ 1852, \aj, 2, 161

\bibitem[Ram{\'{\i}}rez et al.(2009)]{ram09} Ram{\'{\i}}rez, I., 
Mel{\'e}ndez, J., \& Asplund, M.\ 2009, \aap, 508, L17

\bibitem[Ram{\'{\i}}rez et al.(2011)]{ram11} Ram{\'{\i}}rez, I., 
Mel{\'e}ndez, J., Cornejo, D., Roederer, I.~U., \& Fish, J.~R.\ 2011, 
\apj, 740, 76

\bibitem[Schuler et al.(2011)]{sch11} Schuler, S.~C., Cunha, K., Smith, 
V.~V., et al.\ 2011, \apjl, 737, L32

\bibitem[Silva Aguirre et al.(2011)]{sil11} Silva Aguirre, V., Chaplin, 
W.~J., Ballot, J., et al.\ 2011, \apjl, 740, L2

\bibitem[Skumanich(1972)]{sku72} Skumanich, A.\ 1972, \apj, 171, 565

\bibitem[Thoul et al.(1994)]{tbl94} Thoul, A.~A., Bahcall, J.~N., \& Loeb, 
A.\ 1994, \apj, 421, 828

\bibitem[Torres(2010)]{tor10} Torres, G.\ 2010, \aj, 140, 1158

\bibitem[Turner et al.(2001)]{tur01} Turner, N.~H., ten Brummelaar, T.~A., 
McAlister, H.~A., et al.\ 2001, \aj, 121, 3254

\bibitem[Valenti \& Fischer(2005)]{vf05} Valenti, J.~A., \& Fischer, 
D.~A.\ 2005, \apjs, 159, 141

\bibitem[van Leeuwen(2007)]{van07} van Leeuwen, F.\ 2007, \aap, 474, 653

\bibitem[Walker et al.(2007)]{wal07} Walker, G.~A.~H., Croll, B., 
Kuschnig, R., et al.\ 2007, \apj, 659, 1611

\bibitem[Wright et al.(2004)]{wri04} Wright, J.~T., Marcy, G.~W., Butler, 
R.~P., \& Vogt, S.~S.\ 2004, \apjs, 152, 261

\end{thebibliography}
\end{document}